%% file: acl_latex.tex
\pdfoutput=1
\documentclass[11pt]{article}

\usepackage[final]{acl}

\usepackage{times}
\usepackage{latexsym}

\usepackage[T1]{fontenc}
\usepackage[utf8]{inputenc}

\usepackage{microtype}

\usepackage{inconsolata}

\usepackage{graphicx}

\usepackage{multirow}
\usepackage{colortbl}
\usepackage{amsmath}
\usepackage{enumitem} % For custom enumeration

\usepackage{graphicx}
\usepackage{subfigure}
\usepackage{booktabs} % for professional tables
\usepackage{amsmath}
\usepackage{natbib}
\usepackage{tcolorbox}
\usepackage{hyperref}       % hyperlinks
\usepackage{url}            % simple URL typesetting
\usepackage{booktabs}       % professional-quality tables
\usepackage{amsfonts}       % blackboard math symbols
\usepackage{nicefrac}       % compact symbols for 1/2, etc.
\usepackage{microtype}      % microtypography
\usepackage{xcolor}         % colors

\usepackage{mathrsfs}
\usepackage{booktabs}

\usepackage{cuted}%%\stripsep-3pt
\usepackage{float}%%%%提供浮动体的[H]选项，进而取消浮动
\usepackage{caption}%%提供\captionof命令

\newtcolorbox{mybox}[1][]{
  arc=1mm,
  boxrule=1pt,
  colback=gray!20, % 更改背景颜色为灰色
  colframe=black!80,
  fonttitle=\bfseries,
  fontupper=\small, % 更改字体大小为小号
  title=#1,
  left=1mm,
  right=1mm,
  top=1mm,
  bottom=1mm
}

\newcommand{\framework}{\textsc{mABC}}

\newcommand{\SmallEquation}{\fontsize{8.0pt}{\baselineskip}\selectfont}

\NewDocumentCommand\emojiowl{}{
$\vcenter{\hbox{\includegraphics[height=1.2em]{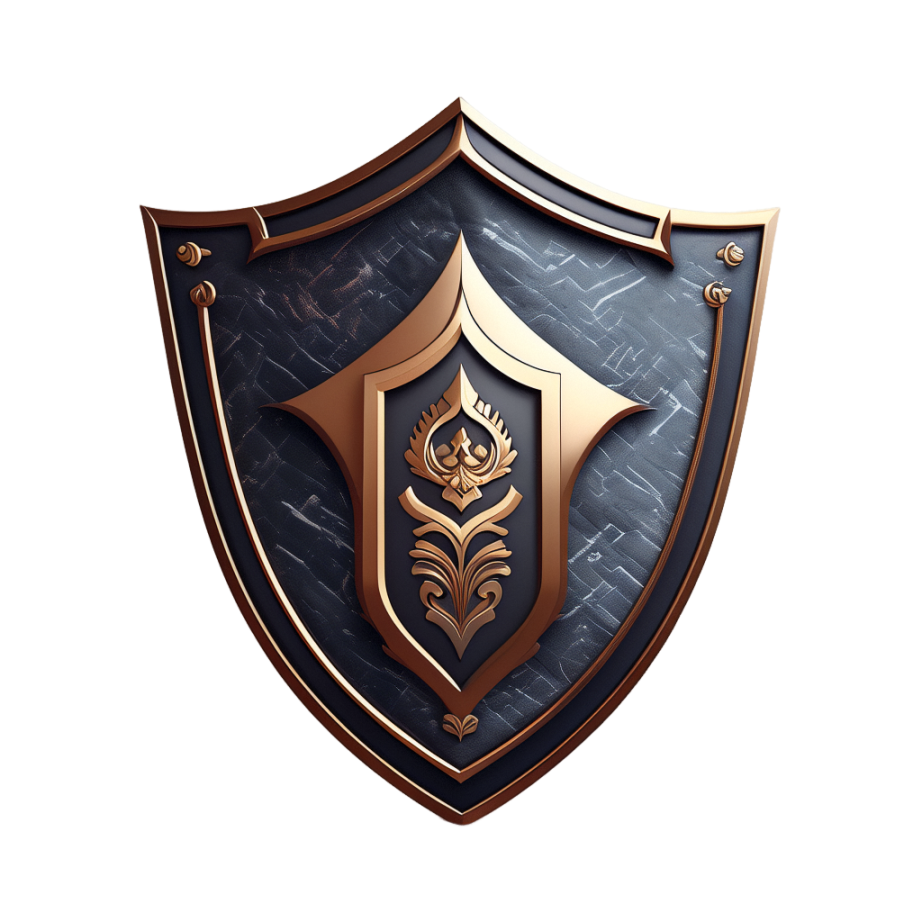}}}$
}

\title{\emojiowl{}\framework{}: Multi-Agent Blockchain-inspired Collaboration for Root Cause Analysis in Micro-Services Architecture}

\author{Wei Zhang\textsuperscript{\textrm{1}}, Hongcheng Guo\textsuperscript{\textrm{1}*}, Jian Yang\textsuperscript{\textrm{1}*}, Zhoujin Tian\textsuperscript{\textrm{1}}, Yi Zhang\textsuperscript{\textrm{1}}, Chaoran Yan\textsuperscript{\textrm{1}}, \\ \textbf{Zhoujun Li\textsuperscript{\textrm{1}*}, Tongliang Li\textsuperscript{\textrm{2}}, Xu Shi\textsuperscript{\textrm{3}}, Liangfan Zheng\textsuperscript{\textrm{3}}, Bo Zhang\textsuperscript{\textrm{3}}} \\
\textsuperscript{\textrm{1}}State Key Laboratory of Complex \& Critical Software Environment, Beihang University \\
\textsuperscript{\textrm{2}}Computer School, Beijing Information Science and Technology University \\
\textsuperscript{\textrm{3}}Cloudwise Research \\
\texttt{\{zwpride,hongchengguo,jiaya,eitbar,zhangyi2021,ycr2345,lizj\}@buaa.edu.cn;} \\
\texttt{tonyliangli@bistu.edu.cn;\{tim.shi,leven.zheng,bowen.zhang\}@cloudwise.com;}\thanks{Corresponding author.}
}

\begin{document}
\maketitle
\begin{abstract}
Root cause analysis (RCA) in Micro-services architecture (MSA) with escalating complexity encounters complex challenges in maintaining system stability and efficiency due to fault propagation and circular dependencies among nodes. Diverse root cause analysis faults require multi-agents with diverse expertise. To mitigate the hallucination problem of large language models (LLMs),  we design blockchain-inspired voting to ensure the reliability of the analysis by using a decentralized decision-making process. To avoid non-terminating loops led by common circular dependency in  MSA, we objectively limit steps and standardize task processing through \textit{Agent Workflow}. We propose a pioneering framework, \textbf{m}ulti-\textbf{A}gent \textbf{B}lockchain-inspired \textbf{C}ollaboration for root cause analysis in micro-services architecture (\framework{}), where multiple agents based on the powerful LLMs follow \textit{Agent Workflow} and collaborate in blockchain-inspired voting. Specifically, seven specialized agents derived from \textit{Agent Workflow} each provide valuable insights towards root cause analysis based on their expertise and the intrinsic software knowledge of LLMs collaborating within a decentralized chain. Our experiments on the AIOps challenge dataset and a newly created Train-Ticket dataset demonstrate superior performance in identifying root causes and generating effective resolutions. The ablation study further highlights \textit{Agent Workflow}, multi-agent, and blockchain-inspired voting is crucial for achieving optimal performance. \framework{} offers a comprehensive automated root cause analysis and resolution in micro-services architecture and significantly improves the IT Operation domain. The code and dataset are in \url{https://github.com/zwpride/mABC}.
\end{abstract}

\input{content/intro}
\input{content/method}

\input{content/experiment}
\input{content/result}
\input{content/related_work}
\input{content/conclusion}
\input{content/limitation}

% \clearpage

\bibliography{custom}

\clearpage
\appendix
\input{content/appendix}

\end{document}

%% file: content/intro.tex
\section{Introduction}
Micro-services architecture (MSA) decomposes an application into a series of independent nodes, interacting through lightweight communication mechanisms \cite{zhang2021cloudrca, kim2013root, alquraan2018analysis}. A key component in maintaining MSA is Root cause analysis (RCA), which aims to find the root cause of alert events and enhance system robustness plays a significant role in avoiding data leaks and program failure analysis \cite{lin2018microscope,ma2020automap}.

\begin{figure}[t]
    \centering
    \includegraphics[width=1.0\linewidth]{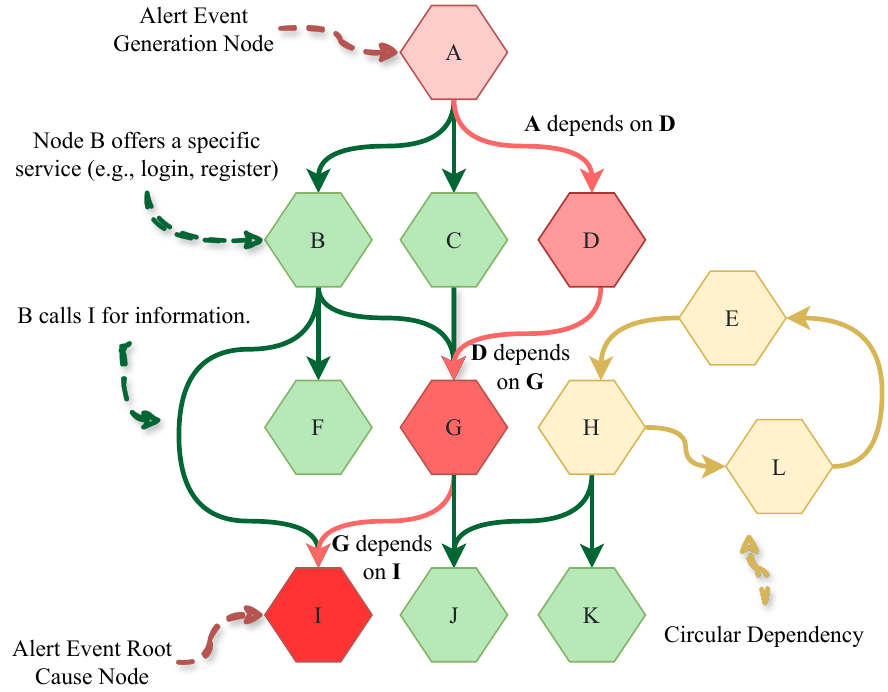}
    \caption{Example of root cause analysis in MSA. Each node corresponds to a specific service in the system (e.g., login, register). Edge \textit{B}$\to$\textit{I} represents that service \textit{I} relies on the information provided by service \textit{B}. Alert event arises on node \textit{A} while alert event root cause node is \textit{I} with fault propagation path \textit{I}$\to$\textit{G}$\to$\textit{D}$\to$\textit{A} where a challenge circular dependency of \textit{H}$\to$\textit{E}$\to$\textit{L}$\to$\textit{H}.}
    \label{fig: intro}
\end{figure}

Compared with traditional architectures only containing one central service, RCA in Micro-services architecture (MSA) has become extremely difficult as faults continue to propagate between service nodes and alerts become increasingly complex~\cite{jamshidi2018microservices, TraceAnomaly}. In Figure~\ref{fig: intro}, alert event arises on \textit{A}, while the alert event root cause node is \textit{I} with fault propagation path \textit{I}$\to$\textit{G}$\to$\textit{D}$\to$\textit{A}. RCA identifies the root cause of faults, \textit{I} here, by tracing back to the origin node and even further analyzing metrics. Existing approaches such as TraceAnomaly \cite{TraceAnomaly}, and MEPFL \cite{zhou2019latent} with lack of mechanism are unable to handle \textbf{circular dependencies} (e.g. \textit{H}$\to$\textit{E}$\to$\textit{L}$\to$\textit{H}) in Figure~\ref{fig: intro} well and rely heavily on supervised training processes. Large language models (LLMs) like GPT \cite{openai2023gpt4} and their integration with multi-agent exhibit remarkable analytical and problem-solving capabilities, which are essential for identifying and addressing the root cause of fault in complex MSA \cite{wei2022finetuned,kojima2022large,wei2022chain,yao2023react}. Although RCA-Copilot \cite{chen2023empowering}, RCAgent \cite{wang2023rcagent}, and D-Bot \cite{zhou2023dbot} have improved RCA tools with event matching and information aggregation, they struggle with \textbf{the hallucination problem} and the common \textbf{cross-node fault} (e.g. \textit{I}$\to$\textit{G}$\to$\textit{D}$\to$\textit{A} in Figure~\ref{fig: intro}) in MSA.

To tackle the above issues, we introduce \framework{}, a groundbreaking framework designed to revolutionize RCA. To solve diverse cross-node faults, we introduce multi-agents with diverse expertise and extensive software knowledge to analyze a wide range of data and navigate through node dependencies, which fully consider the propagation of failures in dependencies. To mitigate hallucination in LLMs, we integrate a blockchain-inspired voting system in the \framework{} that uses multi-agent collaboration and community voting for high-quality content assurance. Inspired by blockchain governance \cite{blockchain, blockchain-wiki}, this process is transparent and community-driven, enhancing decision-making correctness through a decentralized structure. The \framework{} employs dynamic weight adaptation for fairness and includes penalties for inactive or inaccurate agents, alongside a cap on weight concentration. This ensures accuracy, fairness, and reliability in content generation through decentralized, professional multi-agent assessments and repeated verifications. To address the non-terminating loop led by circular dependency, we objectively limit the number of steps and standardize task processing through \textit{Agent Workflow} based on task difficulty and dynamic context perception. By harnessing the power of LLMs within multi-agent blockchain-inspired collaboration, \framework{} conducts RCA and resolution development in MSA, unlike previous methods. Specifically, 1) An alert event arises due to access function blockages or monitoring system alarms in MSA. 2) \textit{Alert Receiver} ($\mathscr{A}_{1}$) chooses and forwards the alert event with the highest priority. 3) \textit{Process Scheduler} ($\mathscr{A}_{2}$) divides unfinished RCA into sub-tasks, handled by \textit{Data Detective} ($\mathscr{A}_{3}$), \textit{Dependency Explorer} ($\mathscr{A}_{4}$), \textit{Probability Oracle} ($\mathscr{A}_{5}$), and \textit{Fault Mapper} ($\mathscr{A}_{6}$) for various requests. 4) \textit{Solution Engineer} ($\mathscr{A}_{7}$) develops resolutions referencing previous successful cases.

Experimental results on the public AIOps challenge dataset and our created train-ticket dataset demonstrate superior performance in identifying root causes and effective resolution development, compared to existing strong baselines. The ablation study further highlights \textit{Agent Workflow}, multi-agent, and blockchain-inspired voting is crucial for achieving optimal performance. Generally, the main contributions of this work are as follows: 

\begin{itemize}
\item \textbf{Multi-Agent Framework in RCA}: Different from previous works designed for single node fault, we proposed framework \framework{} driven by LLM and multi-agent collaboration standardized by \textit{Agent Workflow} performs RCA and resolution development in complex MSA scenarios, which will be open-sourced first.

\item \textbf{Blockchain-Inspired Voting}: By employing blockchain-inspired voting with multi-agent collaboration in \framework{}, dynamically adjusted weights based on \textit{contribution index} and \textit{expertise index} of agents ensure the accuracy and reliability of content generation.

\item \textbf{Impressive Evaluation}: Superior performance in identifying root causes and resolution development effectively both on the public AIOps challenge dataset and our created Train-Ticket dataset. 
\end{itemize}

%% file: content/method.tex
\section{Methodology}

\begin{figure*}[t]
    \centering
    \includegraphics[width=1.0\textwidth]{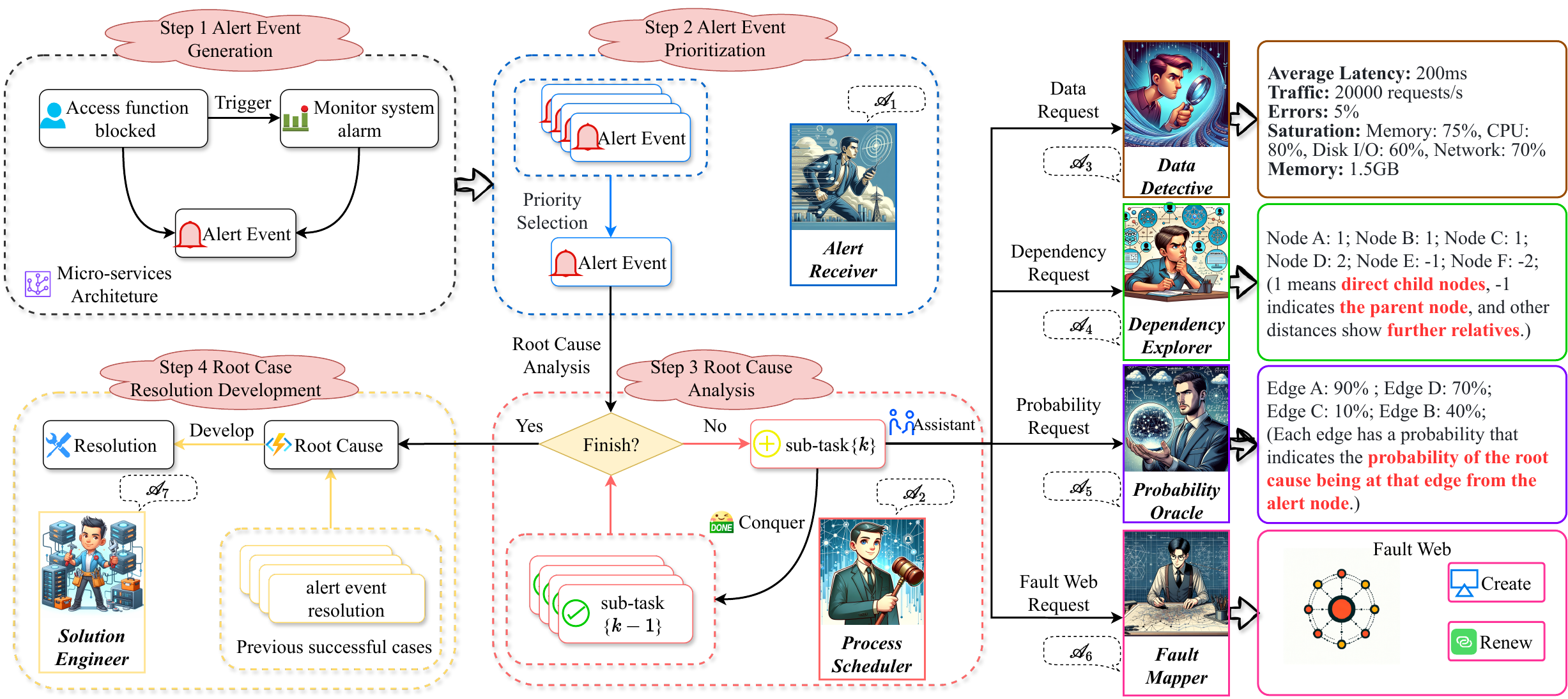}
    \caption{Overview of \framework{}. Overall pipeline encapsulates the flow from alert inception to root cause analysis within \framework{}. 1) An alert event arises due to access function blockages or monitoring system alarms in MSA. 2) \textit{Alert Receiver} ($\mathscr{A}_{1}$) forwards and chooses the alert event with the highest priority. 3) \textit{Process Scheduler} ($\mathscr{A}_{2}$) divides unfinished root cause analyses into sub-tasks, handled by \textit{Data Detective} ($\mathscr{A}_{3}$), \textit{Dependency Explorer} ($\mathscr{A}_{4}$), \textit{Probability Oracle} ($\mathscr{A}_{5}$), and \textit{Fault Mapper} ($\mathscr{A}_{6}$) for various requests. 4) \textit{Solution Engineer} ($\mathscr{A}_{7}$) develops resolutions for the root cause referencing previous successful cases.}
    \label{fig: overview}
    % \vspace{-15pt}
\end{figure*}

\subsection{Overview}
In this section, we provide an overview of \framework{}, specifically engineered to pinpoint the root causes of alert events in a complex MSA. Illustrated in Figure~\ref{fig: overview}, \framework{} introduces seven agents: \textit{Alert Receiver}, \textit{Process Scheduler}, \textit{Data Detective}, \textit{Dependency Explorer}, \textit{Probability Oracle}, \textit{Fault Mapper}, and \textit{Solution Engineer}. These agents collaborate transparently and equally, invoking each other to address alert events in the \framework{} pipeline. In MSA, alert events can arise from user-side blocked function access and monitoring system alarms, such as increased login response times and network latency in the login node. The specific case is shown in Figure~\ref{fig: promptA1},~\ref{fig: promptA2},~\ref{fig: promptA3},~\ref{fig: promptA4},~\ref{fig: promptA5},~\ref{fig: promptA6},~\ref{fig: promptA7} in Appendix~\ref{case}.

\subsection{Agent Workflow}
\begin{figure}[t]
    \centering
    \includegraphics[width=0.9\linewidth]{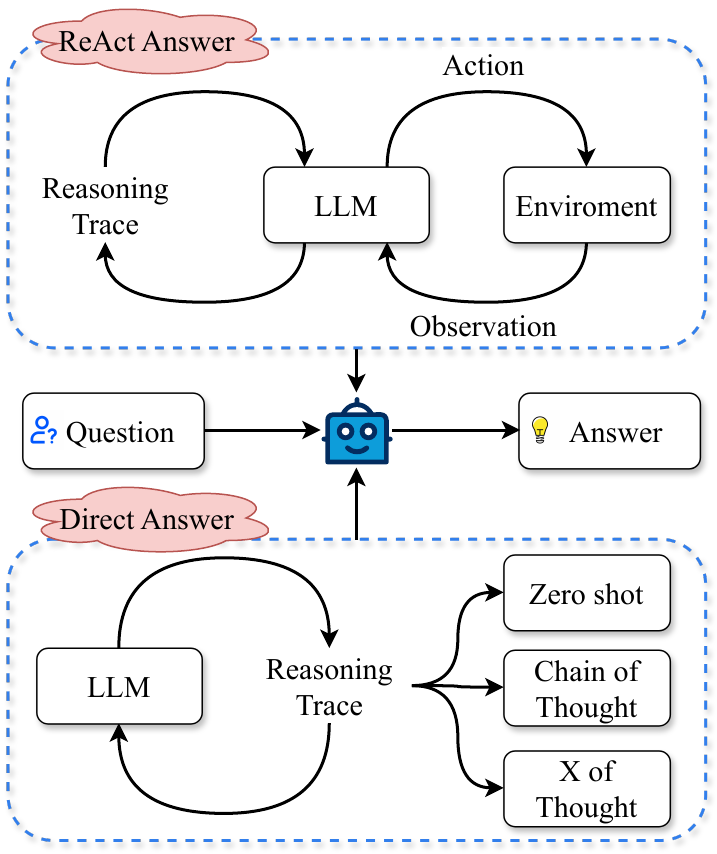}
    \caption{Two distinct workflows of agent.}
    \label{fig: workflow}
    % \vspace{-15pt}
\end{figure}

In Figure~\ref{fig: workflow}, \textit{Agent Workflow} enables all agents to complete their tasks effectively, adhering to a prescribed methodology. For questions that require real-time data or additional information, \textit{Agent Workflow} activates the ReAct answer workflow, which involves an iterative cycle of thought, action, and observation until a satisfactory answer is reached. Conversely, when no external tools are necessary, \textit{Agent Workflow} defaults to the direct answer workflow, where responses are directly formulated based on the prompt provided. It is worth noting that to address the non-terminating loop led by circular dependency, we terminate the process at $20$ steps. The prompt example is shown in Figure~\ref{fig: promptB1},~\ref{fig: promptB2} in Appendix~\ref{workflow}.

\subsection{Multi-Agent}
% 标注agent A-BCD->G, X 属于 A=>G
In this section, we provide a thorough introduction of agents in \framework{}. The role description are shown in Figure~\ref{fig: promptC1},~\ref{fig: promptC2},~\ref{fig: promptC3},~\ref{fig: promptC4},~\ref{fig: promptC5},~\ref{fig: promptC6},~\ref{fig: promptC7} in Appendix~\ref{role and tool} and tools are shown in Table~\ref{table: tools} in Appendix~\ref{tool}.

\subsubsection{Alert Receiver ($\mathscr{A}_{1}$)}
In Figure~\ref{fig: overview}, the responsibility of \textit{Alert Receiver} is to sort the received alert events based on the time, urgency, and scope of impact. After determining the priority of the alert events, \textit{Alert Receiver} dispatches the most urgent and widely impacting alert events to \textit{Process scheduler} further processing following the pipeline.

\subsubsection{Process Scheduler ($\mathscr{A}_{2}$)}
In Figure~\ref{fig: overview}, when an alert arrives at \textit{Alert Receiver}, \textit{Process Scheduler} engages specialized agents for tasks like data gathering, fault web updates, dependency analysis, and probability scoring. It forwards critical insights to \textit{Solution Engineer} for resolution. After each sub-task, it checks for root cause identification. If unresolved, it iterates by generating new sub-tasks and seeking further agent assistance until the root cause is determined. Finally, \textit{Process Scheduler} provides the root cause, an updated fault web, and some resolutions, concluding handling process and preparing for the next alert.

\subsubsection{Data Detective ($\mathscr{A}_{3}$)}
In Figure~\ref{fig: overview}, \textit{Data Detective} collects data from designated nodes within specified time windows as directed by the \textit{Process Scheduler}. To ensure thorough analysis and maximize informational value, it excludes non-essential data and processes key metrics like average latency, traffic volume, error rates, resource saturation, and concurrent user counts into charts and reports. This approach simplifies data handling for LLMs, streamlining the task of \textit{Data Detective} and enhancing efficiency in data exploration and analysis within \framework{}.

\begin{table*}[ht]
\begin{center}
\resizebox{1\textwidth}{!}{
 \small
\begin{tabular}{ll}
   \toprule
   \textbf{Alert Events} & \textbf{Description} \\
   \midrule
Tablespace High Utilization & Indicates \textcolor{red}{extensive data occupation} in tablespace, potentially degrading database performance. \\
   \midrule
Database Connectivity Fault & Signifies possible connection issues due to \textcolor{red}{excessive connections}, impacting response and transactions. \\
   \midrule
CPU Resource Insufficiency & High session average CPU time suggest \textcolor{red}{significant CPU occupation}, risking performance and stability. \\
   \midrule
Memory Overflow & Shows \textcolor{red}{memory usage exceeding safe limits}, risking performance degradation or crashes. \\
   \midrule
Disk IO Performance Fault & \textcolor{red}{Abnormal increase in physical read rates}, indicating potential disk IO issues. \\
\bottomrule
\end{tabular}
}
\caption{Examples of Alert Events}
\label{tab: Examples of Alert Events}
\vspace{-15pt}
\end{center}
\end{table*}

\subsubsection{Dependency Explorer ($\mathscr{A}_{4}$)}
In Figure~\ref{fig: overview}, \textit{Process Scheduler} sends a \textit{Dependency Request} to \textit{Dependency Explorer} to query dependencies among micro-services nodes, including the specific node and alert time. \textit{Dependency Explorer} identifies direct and indirect dependencies based on global topology and calls within the time window. This is crucial for tracing fault paths, marking impacted nodes, and facilitating further root cause analysis and resolution.

\subsubsection{Probability Oracle ($\mathscr{A}_{5}$)}
In Figure~\ref{fig: overview}, \textit{Probability Oracle} assesses the failure probability of nodes. Inaccessible nodes get a high default failure probability, while accessible nodes are evaluated based on performance metrics like response time, error rate using a computational model. By analyzing data correlations, such as a high Pearson correlation coefficient indicating a link between response time and error rate, \textit{Probability Oracle} adjusts failure probabilities of correlated nodes increasingly while decreasing on other nodes. These probabilities are sent to \textit{Process Scheduler}, aiding in updating fault web, root cause analysis, and resolution development.

\subsubsection{Fault Mapper ($\mathscr{A}_{6}$)}
In Figure~\ref{fig: overview}, when \textit{Fault Web} needs to be updated, \textit{Process Scheduler} issues a \textit{Fault Web Request}, which includes nodes and their corresponding fault probabilities. \textit{Fault Mapper} creates or renews \textit{Fault Web} based on this information to visually represent the fault probabilities between different nodes. \textit{Fault Web} not only displays the alert source node but also depicts other related nodes and the fault probabilities of their connecting edges. \textit{Fault Mapper} ensures that \textit{Process Scheduler} can make decisions based on the most up-to-date information, thereby guiding \textit{Solution Engineer} to develop appropriate resolutions.

\subsubsection{Solution Engineer ($\mathscr{A}_{7}$)}
In Figure~\ref{fig: overview}, \textit{Solution Engineer} receives \textit{Root Cause Analysis and Solution Requests} from \textit{Process Scheduler} and then decides the final root cause analysis and development of solutions based on the available node data. \textit{Solution Engineer} performs node-level analysis to confirm the nodes affected to be repaired by the MCA when node downtime data is unavailable. If node data is available, \textit{Solution Engineer} performs metric-level analysis to find the real problem metric through the correlation between metrics and historical value fluctuations to develop more reasonable solutions like increasing disk throughput for high read and write latency. \textit{Solution Engineer} also references previous successful cases, like in Table~\ref{tab: Examples of Alert Events}, to guide the development of the current solution and the conclusion of the process ensuring that the proposed resolution is practical and effective.

\subsection{Blockchain-Inspired Voting}
\subsubsection{Blockchain Communication}
To mitigate the hallucination of LLM and avoid falling into non-terminating loops, we have designed \textit{blockchain-inspired voting} as a reflection for any answer to any question from any agent. After the agent answers, all other agents decide whether to initiate a poll and obtain the result through weighted voting. Answers that do not initiate a poll process or pass in the poll are considered to be of high quality due to the majority approval of the agents, while answers that fail to pass will be regenerated by the author agent to improve the quality. The agents in the \framework{} are transparent and equal to each other, despite their different responsibilities, and compose a decentralized structure \textit{Agent Chain}. Additionally, although \textit{Agent Chain} lacks the implementation of a Byzantine fault-tolerant system, it is still very robust for driven by \textit{Agent Workflow} to avoid the generation of false messages. Inspired by the governance guidelines of decentralized best practice blockchain, we choose on-chain governance to allow participants to trust each other and leave decision-making power in the hands of decentralized entities. More detailed rules description is shown Figure~\ref{fig: promptD1} Appendix~\ref{voting}.
\subsubsection{Voting Weights}
Voting weight is determined by \( w_c \cdot w_e \), where \textit{contribution index} (\( w_c \)) reflects activity level, and \textit{expertise index} (\( w_e \)) reflects professionalism.

\noindent{}The \textit{contribution index} \( w_c \) is updated as follows:
{\SmallEquation
\begin{align}
w_c = \min\left(w_c \cdot (1 - \delta) + \Delta w_c, \; w_{c_{\max}}\right)
\end{align}
}were \( w_c \) starts at 1.0. The decay rate \( \delta \) ranges from 0 to 0.03, applied after each voting event to encourage ongoing contribution and prevent power concentration. \( \Delta w_c \) is an increase of 0.1 from voting participation and proposal submission. \( w_{c_{\max}} \) is set to 1.5 to ensure fairness.

\noindent{}The \textit{expertise index} \( w_e \) is governed by:
{\SmallEquation
\begin{align}
w_e = \min\left(w_e + \Delta w_e, \; w_{e_{\max}}\right)
\end{align}
}where \( w_e \) does not decay automatically, reflecting accumulated expertise. \( \Delta w_e \) increases by 0.01 if the agent's vote aligns with the final outcome and decreases by 0.01 otherwise. \( w_{e_{\max}} \) is also set to 1.5 to prevent disproportionate influence.

The voting weight system balances activity and expertise to ensure fairness. The contribution index (\(w_c\)) starts at 1.0, increases by 0.1 for each vote or proposal, and decays by up to 0.03 after each voting event to encourage ongoing participation, capped at 1.5. The expertise index (\(w_e\)) increases by 0.01 if an agent's vote aligns with the outcome and decreases by 0.01 otherwise, capped at 1.5, reflecting professionalism without decay. This system rewards both active engagement and accurate contributions, preventing power hoarding and reckless voting, while maintaining a balanced and fair decision-making process.

\subsubsection{Voting Outcome Determination} The support rate ($s$) and participation rate ($p$) are defined as:
{\SmallEquation
\begin{align}
s &= \frac{\sum_{i=1}^{n} \mathbf{1}(w_i)}{\sum_{i=1}^{n} w_i} \\
p &= \frac{\sum_{i=1}^{n} \mathbf{1'}(w_i)}{\sum_{i=1}^{n} w_i}
\end{align}
}where $n$ is the total number of voting agents, $\text{vote}_i$ is the vote of the $i$-th agent, and $w_i$ is the weight of the $i$-th vote. A proposal passes if $s \geq \alpha \And p \geq \beta$, where $\alpha$ and $\beta$ are predefined thresholds (e.g., $0.5$). The indicator function $\mathbf{1}(\cdot)$ outputs $w_i$ if the $i$-th agent votes \textbf{For}, and $0$ otherwise. The indicator function $\mathbf{1'}(\cdot)$ outputs $w_i$ if the $i$-th agent votes \textbf{For} or \textbf{Against}, and $0$ for \textbf{Abstain}.

\subsubsection{Voting Process}
\begin{figure}[h]
    \centering
    \includegraphics[width=1.0\linewidth]{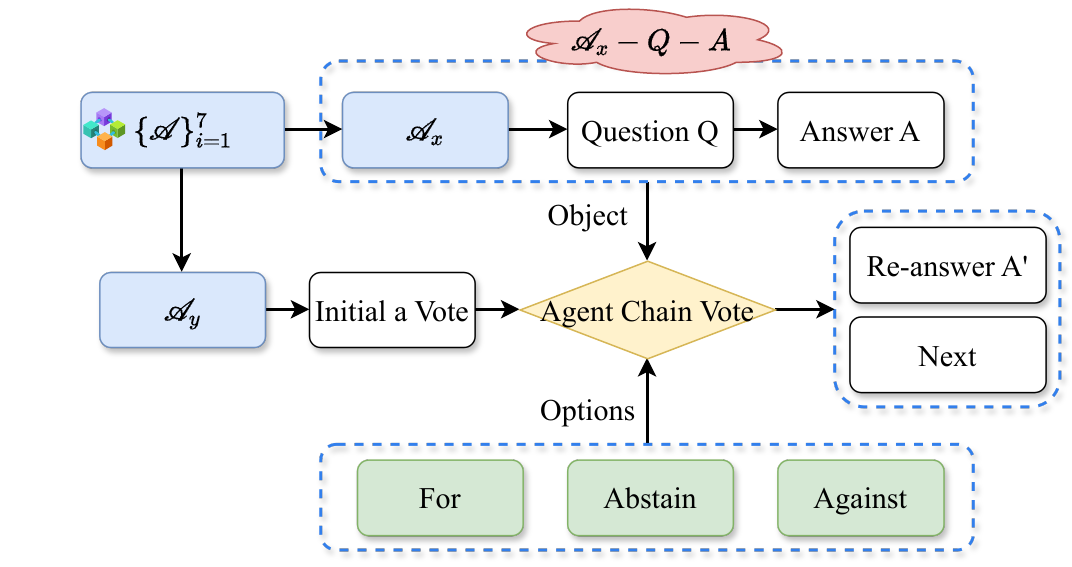}
    \caption{Vote process on \textit{Agent Chain}}
    \label{fig: vote}
\end{figure}

On \textit{Agent Chain}, every agent is entitled to participate in voting. The voting process works in Figure~\ref{fig: vote}: When $\mathscr{A}_{x} \in \{\mathscr{A}_{i}\}_{i=1}^{7}$ gets an answer $A$ for question $Q$, all agents on the chain will examine $A$ and face the choice of whether to initiate a vote on $X-Q-A$. If no Agent initiates a vote, the answer is accepted. If $\mathscr{A}_{y} \in \{\mathscr{A}_{i}\}_{i=1}^{7}$ requires a vote, all agents on the agent chain will vote on $\mathscr{A}_{y}-\mathscr{A}_{x}-Q-A$, with the voting options being \textbf{For}, \textbf{Abstain}, and \textbf{Against}. If the vote passes, $\mathscr{A}_{x}$ will re-answer Question $Q$ to generate a new Answer $A'$. More detailed description and case are shown in ~\ref{fig: promptD2},~\ref{fig: promptD3},~\ref{fig: promptD4} Appendix~\ref{voting}.

%% file: content/experiment.tex
\begin{table*}
  \centering
\resizebox{1.0\textwidth}{!}{
\begin{tabular}{ll}
   \toprule
   \textbf{Process} & \textbf{Description} \\
   \midrule
   Admin Operations & \textcolor{red}{Admin login, site, and route addition}, train information addition, user and contact addition, multiple queries (routes, trains, sites, etc.), updates, deletions, repeat queries. \\
   \midrule
   Normal Flow & \textcolor{red}{User registration and login}, ticket availability search, ticket booking, order status refresh, order payment, and ticket check-in. \\
   \midrule
   Re-book Flow & Registration and login, availability search, booking, latest order status \textcolor{red}{refresh, re-booking, new order payment} (if applicable), check-in. \\
   \midrule
   Re-Book Fail Flow & Registration and login, availability search and booking, order status refresh, successful first re-booking, order payment, \textcolor{red}{failed second re-booking attempt}. \\
   \midrule
   Search Fail to Add & User registration and login, \textcolor{red}{failed ticket search (due to missing stations), admin adds missing info, ticket research and booking, latest order status refresh}. \\
   \midrule
   Consign Preserve & User registration and login, ticket search and booking, order status refresh, order payment, \textcolor{red}{luggage consignment addition, and check-in}. \\
   \midrule
   Preserve Successfully & User registration and login, \textcolor{red}{ticket availability search and booking, order status refresh, payment, and check-in}. \\
   \bottomrule
\end{tabular}
}
\caption{Train-Ticket Process Descriptions}
\label{tab: Train-Ticket Process Descriptions}
\end{table*}

\section{Experiments}
 
\subsection{Datasets}

\paragraph{Train-Ticket Dataset.}
We curate our dataset on Train-Ticket \cite{zhou2018fault,li2022enjoy}, an open-source MSA from Fudan University. We designed 7 processes and 100 virtual users to simulate real operations. Table~\ref{tab: Train-Ticket Process Descriptions} details each process, which includes operations like registration, login, querying, booking, and ticket changes. Users randomly select processes to cover various scenarios. Specifically, we introduce faults by ChaosBlade \cite{ali2021chaosblade} into the system as outlined in Table~\ref{tab: Fault Types}. Please refer to Appendix~\ref{appendix: Train-Ticket} for more details. 

\begin{table}[h]
\begin{center}
\resizebox{1.0\columnwidth}{!}{
 \small
\begin{tabular}{ll}
   \toprule
   \textbf{Category} & \textbf{Case Examples} \\
   \midrule
   \multirow{2}{*}{\centering Network}
   & Packet loss, Frequent retransmission, DNS failures, \\ & bandwidth saturation, high TCP connection setup delays \\
   \midrule
   \multirow{1}{*}{Storage}
   & High I/O latency \\
   \midrule
   \multirow{1}{*}{CPU}
   & High CPU usage by code, CPU frequently grabs \\
   \midrule
   \multirow{1}{*}{Memory}
    & High frequency of FULL GC, memory frequently grabs \\
   \midrule
   \multirow{2}{*}{Code}
   & Exceptions thrown by error codes, HTTP requests,  \\ & returning error codes \\
   \bottomrule
\end{tabular}
}
\caption{Types of Faults Injected in the Experiment}
\label{tab: Fault Types}
\end{center}
\end{table}

% 1. admin_operations 流程
% 登录管理员账户。
% 添加新站点和路线信息。
% 添加车次信息。
% 添加用户和用户联系人。
% 进行多项查询操作，包括所有路线、配置、列车、站点、联系人、订单、价格、所有行程、所有用户信息的查询。
% 更新站点、路线、车次、用户、用户联系人信息。
% 删除之前添加的站点、路线、车次、用户、用户联系人信息。
% 再次执行多项查询操作。
% 2. normal_flow 流程（未直接提供，假定为常规操作流程）
% 用户注册并登录。
% 用户查询列车余票。
% 用户预订车票。
% 用户刷新订单查看状态。
% 用户支付订单。
% 用户检票进站。
% 3. rebook_flow 流程
% 用户注册并登录。
% 用户查询列车余票。
% 用户预订车票。
% 用户刷新订单以获取最新订单状态。
% 用户改签车票到新的行程。
% 用户支付新的订单（如果产生新订单）。
% 用户检票进站。
% 4. rebook_fail_flow 流程
% 用户注册并登录。
% 用户查询列车余票并预订车票。
% 用户刷新订单以获取最新订单状态。
% 用户第一次尝试改签成功。
% 用户支付订单。
% 用户尝试第二次改签，但失败。
% 5. search_fail_add 流程
% 用户注册并登录。
% 用户尝试查询列车余票，但查询失败（由于不存在的起始站或终止站）。
% 管理员登录并添加缺失的站点、路线和车次。
% 用户重新查询列车余票并预订车票。
% 用户刷新订单以获取最新订单状态。
% 6. consign_preserve 流程
% 用户注册并登录。
% 用户查询列车余票并预订车票。
% 用户刷新订单以获取最新订单状态。
% 用户支付订单。
% 用户为订单添加行李寄送服务。
% 用户检票进站。
% 7. preserve_successfully 流程
% 用户注册并登录。
% 用户查询列车余票并预订车票。
% 用户刷新订单以获取最新订单状态。
% 用户支付订单。
% 用户检票进站。

\paragraph{AIOps Challenge Dataset.}
\textit{2020 AIOps International Challenge Dataset} aims to discover alert events and their root causes in micro-service applications, such as cloud platform services, which include containers, service meshes, micro-service, and variable infrastructures. More details are in Appendix~\ref{appendix:AIops}.

\begin{table*}[h!]
\begin{center}
\resizebox{1.0\textwidth}{!}{
 \small
\begin{tabular}{cccccccccc}
  \toprule
  \multirow{2}*{Model} & \multirow{2}*{Base} & \multicolumn{3}{c}{Train-Ticket} &\multicolumn{3}{c}{AIOps} & \multirow{2}*{Average}\\ 
  \cmidrule(lr){3-5}\cmidrule(lr){6-8} & & RA & PA & Average & RA & PA & Average \\ 
  \midrule
Decision Tree & - & 36.8 & 34.7 & 35.8 & 28.3 & 26.7 & 27.5 & 31.6\\
TraceAnomaly & - & 25.3 & 23.5 & 24.4 & 20.1 & 18.9 & 19.5 & 22.0\\
MEPFL & - & 30.3 & 29.1 & 29.7 & 33.7 & 29.7 & 31.7 & 30.7\\
ReAct & GPT-3.5-Turbo & 31.8 & 26.8 & 29.3 & 25.1 & 22.7 & 23.9 & 26.6 \\ 
ReAct & GPT-4-Turbo & 43.0 & 38.9 & 41.0 & 37.5 & 34.4 & 36.0 & 38.5 \\ 
\framework{} & Llama-3-8B-Instruct & 46.1 & 40.9 & 43.5 & 43.0 & 39.9 & 41.5 & 42.5 \\ 
\framework{} & GPT-3.5-Turbo & 48.1 & 42.8 & 45.5 & 41.1 & 36.7 & 38.9 & 42.2 \\ 
\rowcolor{gray!50} \framework{} & GPT-4-Turbo & \textbf{54.4} & \textbf{48.2} &  \textbf{51.3} & \textbf{45.5} & \textbf{39.3} & \textbf{42.4} & \textbf{46.9}\\ 
  \bottomrule
\end{tabular}
}
\caption{Main Results On Train-Ticket Dataset and AIOps challenge Dataset}
\label{tab: Main Result}
\end{center}
\end{table*}

\subsection{Evaluation Metrics}
\paragraph{Root Cause Result Accuracy (RA)}: Following the previous work \cite{liu2023opseval,zhou2023dbot}, we use result accuracy (RA) to quantify the precision of \framework{} in finding the root cause.
\begin{SmallEquation}
\begin{align}
\begin{split}
    \text{RA} = \frac{A_c - \sigma \cdot A_i}{A_t}
\end{split}
\end{align}
\end{SmallEquation}where $A_c$ denotes the number of correct causes, $A_t$ denotes the total number of causes, $A_i$ denotes the number of wrongly detected causes, and $\sigma$ is a hyper-parameter with 0.1 as the default value because we recognize {\it redundant causes is less harmful than missing causes}. Therefore, we limit the identification to a maximum of 4 root causes for an anomaly.

\paragraph{Root Cause Path Accuracy (PA)}: We use the root cause path accuracy (PA) metric, which aims to measure the effectiveness of \framework{} in tracing the correct path from the symptoms (alerts) back to the root causes. The formula for PA similar to RA focuses on path accuracy as:
\begin{SmallEquation}
\begin{align}
\begin{split}
    \text{PA} = \frac{P_c - \tau \cdot P_i}{P_t}
\end{split}
\end{align}
\end{SmallEquation}where $P_c$ denotes the number of correctly identified paths leading to the root cause, $P_t$ is the total number of actual root cause paths present, $P_i$ denotes the number of incorrectly inferred paths, which do not align with the actual root cause paths, and $\tau$ is a hyper-parameter designed to penalize the inaccuracies in path inference, with a default value of 0.2, reflecting the understanding that inaccurately inferred paths are less detrimental than completely missing the correct paths, but there is a stronger emphasis on precision due to the potential complexity and relevance of paths.

\subsection{Baselines}
We choose decision tree \cite{abdallah2018fault}, TraceAnomaly \cite{TraceAnomaly} and MEPFL \cite{zhou2019latent} as unsupervised baselines to compare \framework{}. For ReAct \cite{yao2023react}, we implement by Langchain \cite{langchain}. It is worth noting that RCACopilot \cite{chen2023empowering} and RCAgent \cite{wang2023rcagent} are not open-source, D-Bot \cite{zhou2023dbot} is not suitable for MSA.

\subsection{Implementation and Configuration} 
We implement \framework{} on Ubuntu 22.04, equipped with an Intel Xeon (R) Gold 6348 CPU @2.60GHz, eight NVIDIA H800 GPUs (80 GB), and 528 GB of memory. The software setup includes NVIDIA-SMI version 535.104.05 and CUDA 12.3. We 
set temperature as $0.6$ for LLMs.

\subsection{Main Results}
Based on the results in Table~\ref{tab: Main Result}, we can see that baselines such as Decision Tree\cite{abdallah2018fault}, TraceAnomaly\cite{TraceAnomaly}, and MEPFL\cite{zhou2019latent} achieved average performance scores ranging from 16.0 to 26.7 on the Train-Ticket dataset and AIOps Challenge dataset. In comparison, ReAct\cite{yao2023react} with GPT-3.5-Turbo and GPT-4-Turbo showed improvements with average scores of 21.6 and 27.5, respectively. However, our proposed \framework{} significantly outperformed all the baseline models and ReAct with GPT-4-Turbo, achieving an impressive average score of 64.9. This indicates that our framework \framework{} has a strong predictive capability and robustness in detecting faults and anomalies in both datasets. The substantial improvement over the baselines demonstrates the effectiveness and superiority of our approach in this context.

%% file: content/result.tex
\section{Analysis}

\subsection{Decision Efficiency.}
Following RCAgent \cite{wang2023rcagent}, we use pass rate (PR) and average path length (APL) to evaluate the thinking trajectory steps of \framework{} in accomplishing the task, considering the validness of action trajectories and stability of the autonomous agent. PR calculated by $\frac{N_p}{N_t}$, where $N_p$ denotes the number of trajectories completed within $\theta$ steps, $\theta$ typically set to $15$, and $N_t$ denotes the total number of trajectories. Besides, APL is denoted by $\frac{\sum_{k=1}^{N_p} L_k}{N_p}$, where $L_k$ denotes the path length of the $k$-th successful trajectory.
\begin{table}[htbp]
\begin{center}
\resizebox{1.0\linewidth}{!}{
\begin{tabular}{ccccccc}
  \toprule
  \multirow{2}*{Model} & \multirow{2}*{Base} & \multicolumn{2}{c}{Train-Ticket} &\multicolumn{2}{c}{AIOps}\\ 
  \cmidrule(lr){3-4}\cmidrule(lr){5-6} & & PR & APL & PR & APL \\ 
  \midrule
Decision Tree & - & 62.4 & 12.1 & 53.8 & 13.4 \\
TraceAnomaly & - & 25.3 & 20.3 & 31.1 & 19.1 \\
MEPFL & - & 33.3 & 19.2 & 37.1 & 18.7 \\
ReAct & GPT-3.5-Turbo & 41.7 & 15.9 & 38.0 & 16.2 \\ 
ReAct & GPT-4-Turbo & 47.1 & 13.9 & 44.2 & 14.3 \\ 
\framework{} & Llama-3-8B-Instruct & 56.1 & 14.8 & 46.1 & 17.7 \\ 
\framework{} & GPT-3.5-Turbo & 58.1 & 13.8 & 51.1 & 14.7 \\ 
\framework{} & GPT-4-Turbo & \textbf{73.0} & \textbf{10.4} & \textbf{68.8} & \textbf{11.7}\\ 
  \bottomrule
\end{tabular}
}
\caption{Decision Efficiency Evaluation}
\label{tab: Decision Efficiency}
\end{center}
\end{table}

\begin{table}[htbp]
\begin{center}
\resizebox{1.0\linewidth}{!}{
 \small
\begin{tabular}{ccccc}
  \toprule
  Model & Base & R-Useful (Train) & R-Useful (AIOps)\\ 
  \midrule
Decision Tree & - & - & - \\
TraceAnomaly & - & - & - \\
MEPFL & - & - & - \\
ReAct & GPT-3.5-Turbo & 2.1 & 2.1\\ 
ReAct & GPT-4-Turbo & 2.4 & 2.3\\ 
\framework{} & Llama-3-8B-Instruct & 3.3 & 2.7 \\ 
\framework{} & GPT-3.5-Turbo & 3.1 & 3.2 \\ 
\framework{} & GPT-4-Turbo & \textbf{4.2} & \textbf{3.6} \\ 
  \bottomrule
\end{tabular}
}
\caption{Human Evaluation}
\label{tab: Human evaluation}
\end{center}
\end{table}

\begin{table*}[h!]
\begin{center}
\resizebox{1.0\linewidth}{!}{
 \small
\begin{tabular}{cccccccccccc}
  \toprule
  \multirow{2}*{Model} &\multicolumn{5}{c}{Train-Ticket} &\multicolumn{5}{c}{AIOps}\\
  \cmidrule(lr){2-6}\cmidrule(lr){7-11} & RA & PA & PR & APL & R-Useful & RA & PA & PR & APL & R-Useful \\
  \midrule

\framework{} & \textbf{54.4} & \textbf{48.2} & \textbf{73.0} & \textbf{10.4} & \textbf{4.2} & \textbf{45.5} & \textbf{39.3} & \textbf{68.8} & \textbf{11.7} & \textbf{3.6} \\ 

\framework{} w/o \textit{Agent Workflow}  & 46.2 & 38.7 & 67.7 & 11.8 & 3.5 & 36.6 & 34.3 & 61.3 & 11.7 & 3.3 \\ 

\framework{} w/o \textit{Multi-Agent} & 38.4 & 33.0 & 52.9 & 13.7 & 2.8 & 32.4 & 28.8 & 50.1 & 13.7 & 2.7 \\ 

\framework{} w/o Voting & 44.8 & 39.9 & 65.7 & 10.9 & 3.3 & 40.1 & 36.7 & 68.0 & 10.2 & 3.4 \\ 
  \bottomrule
\end{tabular}
}
\caption{Component Impact Evaluation}
\label{tab: Component Imapct}
\end{center}
\end{table*}

In Table ~\ref{tab: Decision Efficiency}, the results confirm that \framework{} exhibits superior decision efficiency, as evidenced by its high pass rate (PR) and low average path length (APL) across both datasets, which demonstrates that \framework{} not only completes tasks with a higher probability but also does so with fewer steps, indicating a more efficient and stable decision-making process. The GPT-4-Turbo variant, in particular, showcases the most effective decision trajectories, suggesting that \framework{} is highly capable of generating efficient actions in the context of root cause analysis in a MSA.

\subsection{Human evaluation.}
In Appendix~\ref{human evaluation}, we evaluated 200 randomly selected cases, focusing on root causes, pathways, and resolutions. Ten AIOps experts rated each case on a scale of 1 (very useless) to 5 (very useful), and we averaged these ratings to derive the \textit{Resolution Evaluation Metrics (R-Useful)} score.

Table ~\ref{tab: Human evaluation} highlights a clear preference among experts for the resolutions generated by the \framework{} model, especially when enhanced with GPT-4-Turbo. The higher R-Useful scores for \framework{} with GPT-4-Turbo across both datasets underscore its ability to produce highly useful solutions that align well with expert expectations in the AIOps domain. In contrast, the moderate R-Useful scores for ReAct indicate its limited effectiveness in meeting the nuanced needs of AIOps experts. Decision tree, TraceAnomaly, and MEPFL are not evaluated for R-Useful due to their inability to generate resolutions. Overall, the human evaluation confirms that \framework{} with GPT-4-Turbo excels in creating expert-aligned solutions, showing significant potential for improving decision-making and productivity in AIOps.

\subsection{Component Impact.}
In this section, we verify the impact of three components in \framework{} with GPT-4-Turbo, i.e., \framework{} without \textit{Agent Workflow} (based on ReAct rather than  \textit{Agent Workflow}), \framework{} without \textit{Multi-Agent} (\textit{Agent Workflow}), and \framework{} without Blockchain-Inspired Voting.

Table~\ref{tab: Component Imapct} shows that \framework{} in its complete form excels across all metrics on both datasets, highlighting the necessity of integrating all components. Removing \textit{Agent Workflow} significantly reduces performance, indicating its crucial role. Limiting the framework to a Single Agent results in the lowest scores, severely diminishing its capability for AIOps tasks. Excluding the Blockchain-Inspired Voting component also decreases performance, though less critically, underscoring its role in refining and validating resolutions. The evaluation underscores the importance of each component: \textit{Agent Workflow} provides a structured approach, the \textit{Multi-Agent} architecture captures diverse perspectives, and the Blockchain-Inspired Voting mechanism ensures consensus and reliability. Together, these components synergize to enhance performance of \framework{}, making it a robust tool for root cause analysis in MSA.

%% file: content/related_work.tex
\section{Related Work}

\subsection{Root Cause Analysis in Micro-Services Architecture}
Root cause analysis (RCA) in large systems, particularly within MSA, is crucial in AIOps \cite{alquraan2018analysis, guo2024logformer, zhang2021understanding, liu2019bugs, lou2020understanding}. RCA tasks focus on logs, metrics, and traces, with various studies proposing methods for each data source \cite{guo2021translog, leesatapornwongsa2017scalability,liu2023opseval}. Techniques include identifying failure patterns \cite{ma2020diagnosing, zhang2021cloudrca} and exploring service dependency graphs using metrics and traces \cite{ma2020automap, li2022mining}. Advanced methodologies, particularly NLP for log analysis and anomaly detection, are emphasized \cite{ghosh2022fight,guo2023loglg}. Machine learning has been leveraged for log analysis \cite{locke2021logassist,guo2021logbert,gao2018empirical}, and LLMs are used to enhance RCA performance \cite{zhang2024lemur}.

\subsection{LLM in Micro-Services Architecture}
The rapid advancements in language modeling, particularly through Transformer-based architectures and LLMs like GPT-4 and PaLM \cite{openai2023gpt4, palm2, owl}, have significantly impacted natural language processing and facilitated their use in complex MSA \cite{scaling_laws_gmlm, kaplan2020scaling, park2023generative}. Integrating LLMs with external tools and APIs enhances their functionality in cloud RCA \cite{qin2023toolllm}, improving log analysis and anomaly detection. LLMs are being explored as core intelligence in autonomous multi-agent systems \cite{wang2024executable, zhang2024eclipse}, enabling effective environment interaction \cite{wei2022chain, ouyang2022training}. This spans tasks from toy examples to real-world cloud RCA, highlighting LLM versatility in dynamic environments \cite{wang2023interactive}. The shift towards LLMs in MSA promotes greater autonomy, intelligence, and efficiency \cite{chen2019understanding, chen2023empowering}, driving the growth of end-to-end intelligent operation and maintenance with tasks like database diagnosis, event processing, and RCA \cite{wang2023rcagent, zhou2023dbot}.

%% file: content/conclusion.tex
\section{Conclusion}
In this paper, we introduce \framework{}, a framework that improves alert events resolution in complex MSA by combining multi-agent systems, LLMs, and blockchain voting. We also develop the train-ticket benchmark, an open-source dataset for RCA in MSA. Experimental results on the AIOps challenge and train-ticket datasets show \framework{}'s effectiveness in identifying root causes and providing solutions, with the \textit{Agent Workflow} and voting mechanism being crucial. \framework{} enhances root cause analysis, boosting system reliability and operational efficiency. Future work will focus on enhancing components, incorporating more data sources, and improving agent collaboration, aiming to make \framework{} essential for IT operations.

%% file: content/limitation.tex
% \clearpage
\section{Limitations}
\framework{} faces challenges in complexity and scalability as the number of agents and alert events increase, leading to higher computational overhead and longer processing times, necessitating more efficient algorithms and optimization techniques. Its effectiveness also heavily relies on the accuracy and reliability of the data and models used, requiring regular updates and validations to prevent erroneous analyses. Additionally, the blockchain-inspired voting mechanism, while innovative, can be cumbersome and time-consuming, especially with frequent alert events and numerous agents, potentially delaying decision-making. Future iterations should refine the voting process and incorporate mechanisms to detect and mitigate potential biases among agents. Addressing these limitations will be crucial for enhancing \framework{}'s overall performance and reliability.

\section{Ethical Considerations}
In our study focused on the \framework{}, we exclusively utilized publicly available data and adhered strictly to ethical and legal standards. Sensitive terminology and methodologies were carefully managed to ensure no breach of privacy or confidentiality. Our commitment to transparency and integrity in handling data ensured that our research remained within ethical boundaries without compromising the effectiveness of our findings.

%% file: content/appendix.tex
\section{Alert Event Case}\label{case}

To clearly and directly demonstrate the process in \framework{} and 
intuitively and distinctly display the root cause analysis in a micro-services architecture, we will show how \framework{} works to handle the case in Figure~\ref{fig: intro}. 

\noindent\framework{} uses the alert event arising on node \textit{A} and trace it back to its root cause \textit{I} with fault propagation path \textit{I}$\to$\textit{G}$\to$\textit{D}$\to$\textit{A}. In this simple case, we only demonstrate the summary question and answer for length limitation when an alert arises on node A will be sent to Alert Receiver (\(\mathscr{A}_{1}\)).

\begin{figure}[h]
    \includegraphics[width=1.0\linewidth]{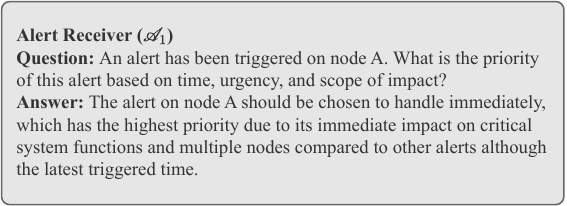}
    \caption{Determining the priority of an alert on node A based on time, urgency, and impact scope for Alert Receiver (\(\mathscr{A}_{1}\)).}
    \label{fig: promptA1}
\end{figure}

\begin{figure}[h]
    \includegraphics[width=1.0\linewidth]{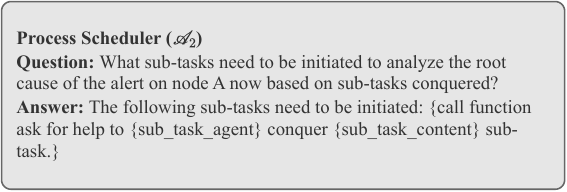}
    \caption{Initiating sub-tasks to analyze the root cause of the alert on node A for Process Scheduler ($\mathscr{A}_{2}$).}
    \label{fig: promptA2}
\end{figure}

\begin{figure}[h]
    \includegraphics[width=1.0\linewidth]{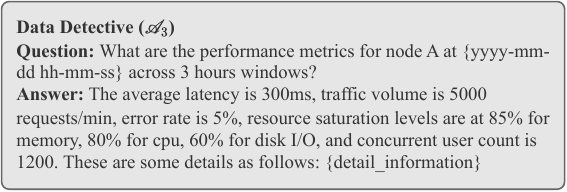}
    \caption{Asking for performance metric of specific node and data collection for Data Detective ($\mathscr{A}_{3}$).}
    \label{fig: promptA3}
\end{figure}
\begin{figure}[h]
    \includegraphics[width=1.0\linewidth]{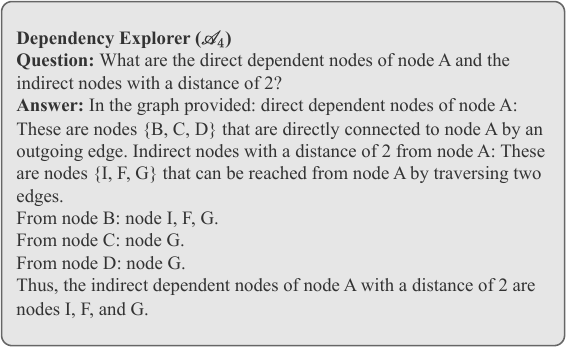}
    \caption{Dependency analysis on specific node for Dependency Explorer ($\mathscr{A}_{4}$).}
    \label{fig: promptA4}
\end{figure}
\begin{figure}[h]
    \includegraphics[width=1.0\linewidth]{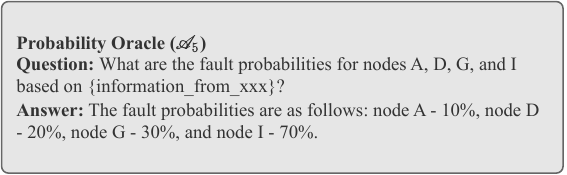}
    \caption{
    Fault probabilities analysis on specific nodes for Probability Oracle ($\mathscr{A}_{5}$).}
    \label{fig: promptA5}
\end{figure}
\begin{figure}[h]
    \includegraphics[width=1.0\linewidth]{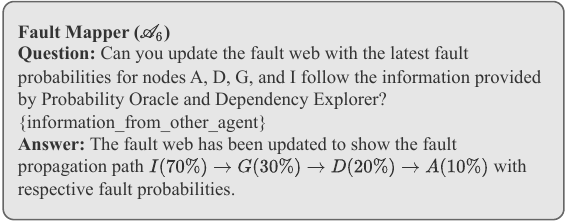}
    \caption{Summarizing information from other agent and update fault web for Fault Mapper ($\mathscr{A}_{6}$).}
    \label{fig: promptA6}
\end{figure}
\begin{figure}[h!]
    \includegraphics[width=1.0\linewidth]{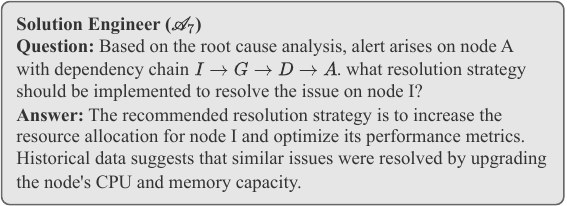}
    \caption{Summarizing the process of root cause analysis and develop a solution to handle the fault for Solution Engineer ($\mathscr{A}_{7}$).}
    \label{fig: promptA7}
\end{figure}

\clearpage
\section{More Details on Train-Ticket} \label{appendix: Train-Ticket}
Train-Ticket is a Kubernetes-deployed train booking system with integrated monitoring and analysis tools, comprising 41 micro-services for high-concurrency functions like ticket query, reservation, payment, changes, and notifications. We create a dataset of 233,111 call chains with 800,656 spans across 112 time periods and 53 nodes, including 900 direct alert events and 294 induced by external nodes.
\begin{figure}[h]
    \centering
    \includegraphics[width=0.8\columnwidth]{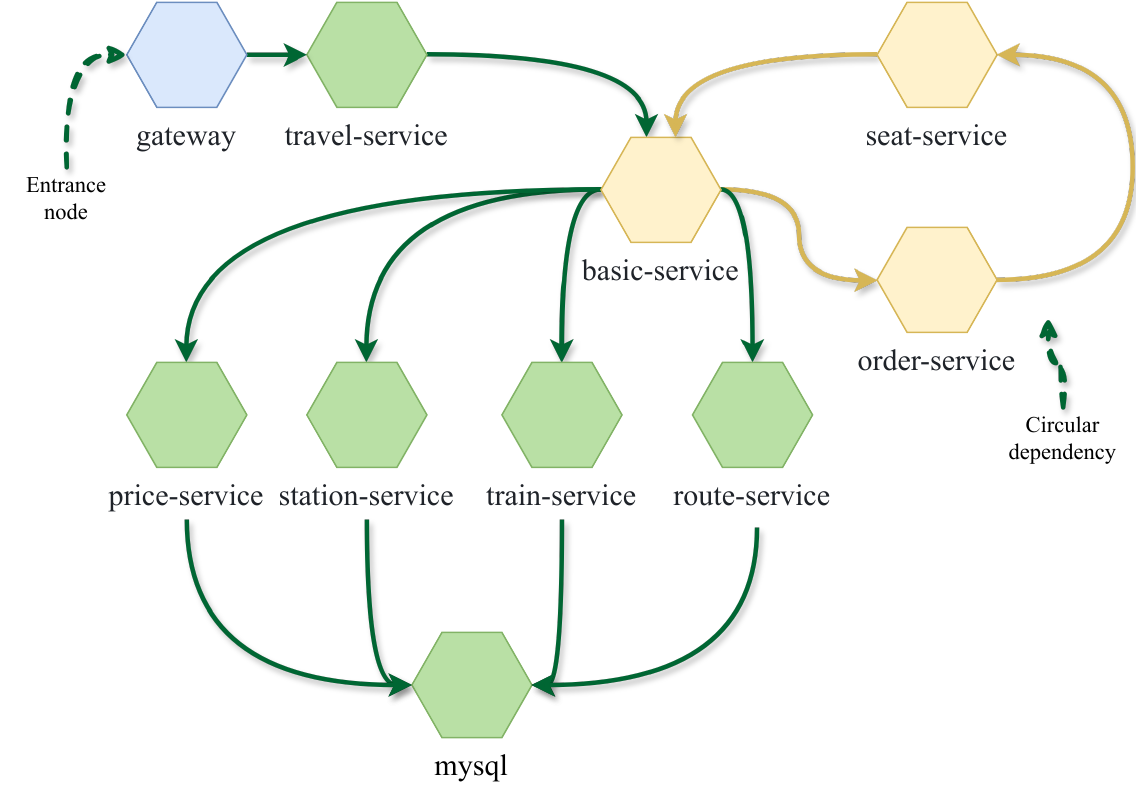}
    \caption{An example of Train-Ticket architecture (query remaining tickets). \textit{basic service} relies on \textit{seat service}, \textit{order service}, and others. A circular dependency of \textit{basic-service}$\to$\textit{order-service}$\to$\textit{seat service}$\to$\textit{basic-service} brings a new challenge for root cause analysis.}
\end{figure}

\section{More Details on AIOps Challenge Dataset} \label{appendix:AIops}
In Table~\ref{tab: AIOpsSUM}, the types of alert events mainly include container CPU utilization, container memory utilization, database connection limit, database close, host network delay, and container network loss. All types of alert events are distributed across various nodes of the system. The dataset includes 14 days of system logs totaling 145,907,050 entries. 
\begin{table}[h]
\begin{center}
\resizebox{0.65\columnwidth}{!}{
 \small
\begin{tabular}{ccc}
   \toprule
   \textbf{Alert Node} & \textbf{Alert Events Type} & \textbf{Count} \\
   \midrule
   os\_021 & CPU & 8434 \\
   docker\_006 & Database Connectivity & 130 \\
   docker\_006 & Database Local Method & 7174 \\
   os\_021 & Operate System & 6352 \\
   docker\_008 & Database Connectivity & 233 \\
   docker\_008 & Database Local Method & 7552 \\
   docker\_005 & Database Connectivity & 211 \\
   docker\_005 & Database Local Method & 9684 \\
   os\_022 & CPU & 15099 \\
   docker\_001 & Network & 2 \\
   os\_022 & Operate System & 109 \\
   docker\_004 & Network & 124 \\
   \bottomrule
\end{tabular}
}
\caption{Summary of Alert Events in 2020 International AIOps Challenge Dataset}
\label{tab: AIOpsSUM}
\end{center}
\end{table}

\section{Prompts for Agent Workflow}\label{workflow}
In this section, we will introduce the workflow prompt for two distinct agent workflows to demonstrate how agents work in their own tasks.
\begin{figure}[h]
    \includegraphics[width=1.0\linewidth]{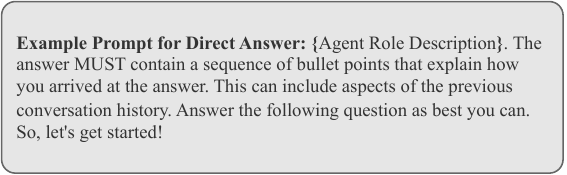}
    \caption{Example Prompt for Direct Answer.}
    \label{fig: promptB1}
\end{figure}
\begin{figure}[h]
    \includegraphics[width=1.0\linewidth]{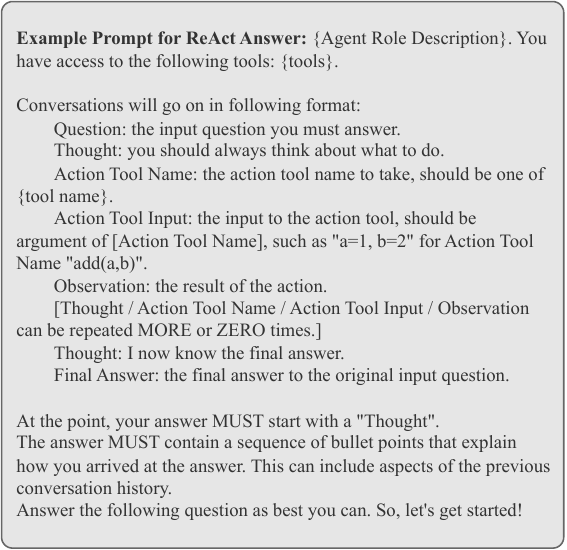}
    \caption{Example Prompt for ReAct Answer.}
    \label{fig: promptB2}
\end{figure}

\section{Agent Role Description Prompt for Multi-Agent}\label{role and tool}
In this section, we will introduce the agent role description prompt for each agent in multi-agent.

\begin{figure}[h]
    \includegraphics[width=1.0\linewidth]{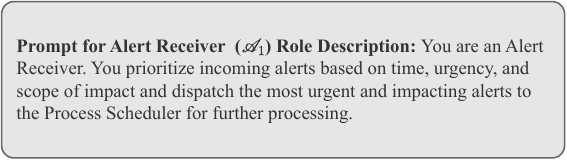}
    \caption{Prompt for Alert Receiver ($\mathscr{A}_{1}$) Role Description.}
    \label{fig: promptC1}
\end{figure}

\begin{figure}[h]
    \includegraphics[width=1.0\linewidth]{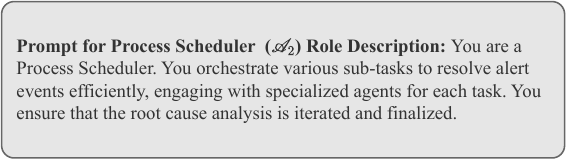}
    \caption{Prompt for Process Scheduler ($\mathscr{A}_{2}$) Role Description.}
    \label{fig: promptC2}
\end{figure}

\begin{figure}[h]
    \includegraphics[width=1.0\linewidth]{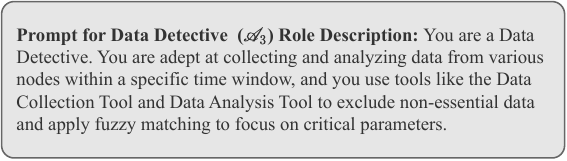}
    \caption{Prompt for Data Detective ($\mathscr{A}_{3}$) Role Description.}
    \label{fig: promptC3}
\end{figure}

\begin{figure}[h]
    \includegraphics[width=1.0\linewidth]{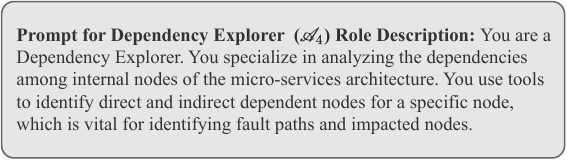}
    \caption{Prompt for Dependency Explorer ($\mathscr{A}_{4}$) Role Description.}
    \label{fig: promptC4}
\end{figure}

\begin{figure}[h]
    \includegraphics[width=1.0\linewidth]{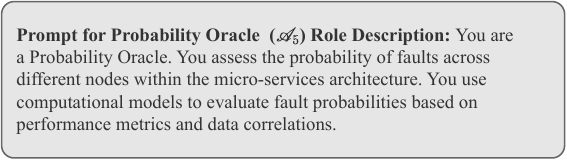}
    \caption{Prompt for Probability Oracle ($\mathscr{A}_{5}$) Role Description.}
    \label{fig: promptC5}
\end{figure}

\begin{figure}[h]
    \includegraphics[width=1.0\linewidth]{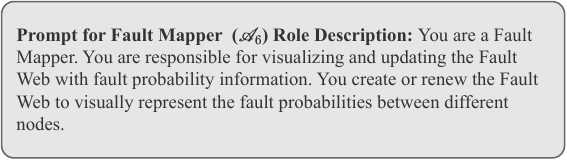}
    \caption{Prompt for Fault Mapper ($\mathscr{A}_{6}$) Role Description.}
    \label{fig: promptC6}
\end{figure}

\begin{figure}[h!]
    \includegraphics[width=1.0\linewidth]{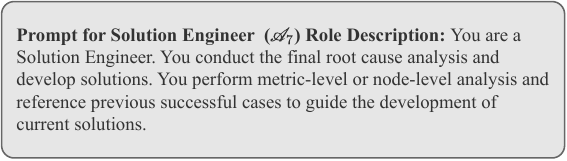}
    \caption{Prompt for Solution Engineer ($\mathscr{A}_{7}$) Role Description.}
    \label{fig: promptC7}
\end{figure}

\clearpage

\section{Blockchain Voting}\label{voting}

% \subsection{Prompt for Rule Description in Blockchain Voting.}
% \subsection{Prompt for Initiating a Poll in Blockchain Voting}
% \subsection{Prompt for Voting in Blockchain Voting}

\begin{figure}[h]
    \includegraphics[width=1.0\linewidth]{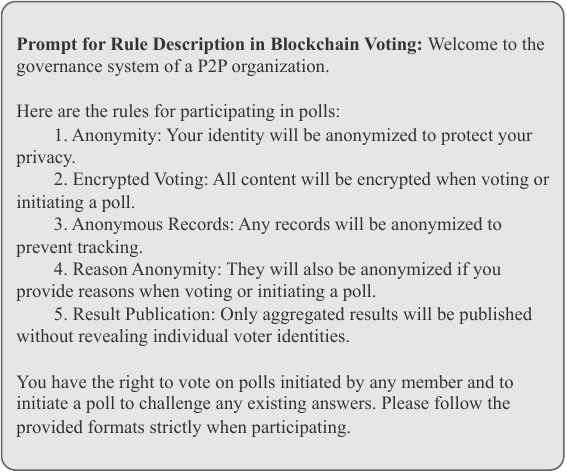}
    \caption{Prompt for Rule Description in Blockchain Voting.}
    \label{fig: promptD1}
\end{figure}
\vspace{100pt}
\begin{figure}[h]
    \includegraphics[width=1.0\linewidth]{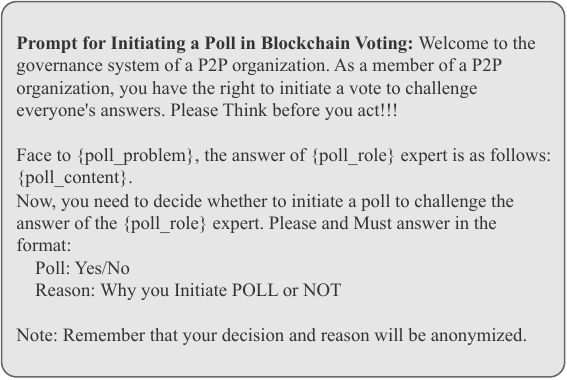}
    \caption{Prompt for Initiating a Poll in Blockchain Voting.}
    \label{fig: promptD2}
\end{figure}

\begin{figure}[h]
    \includegraphics[width=1.0\linewidth]{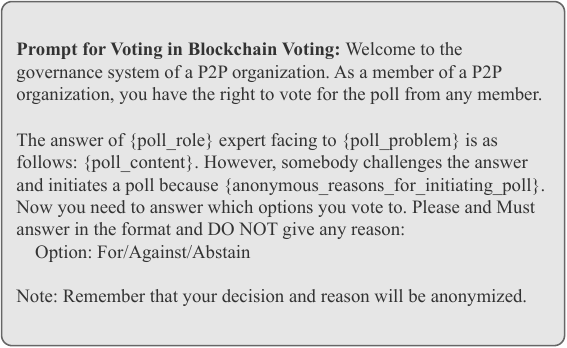}
    \caption{Prompt for Voting in Blockchain Voting.}
    \label{fig: promptD3}
\end{figure}

\begin{figure}[h]
    \includegraphics[width=1.0\linewidth]{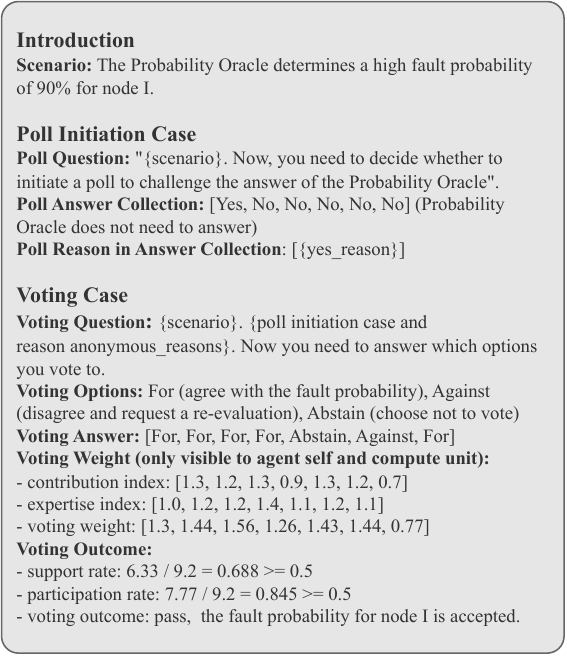}
    \caption{Case in Blockchain Voting by \framework{}.}
    \label{fig: promptD4}
\end{figure}

\clearpage

\section{Human Evaluation}\label{human evaluation}
\begin{figure}[!h]
    \centering
    \includegraphics[width=2.0\linewidth]{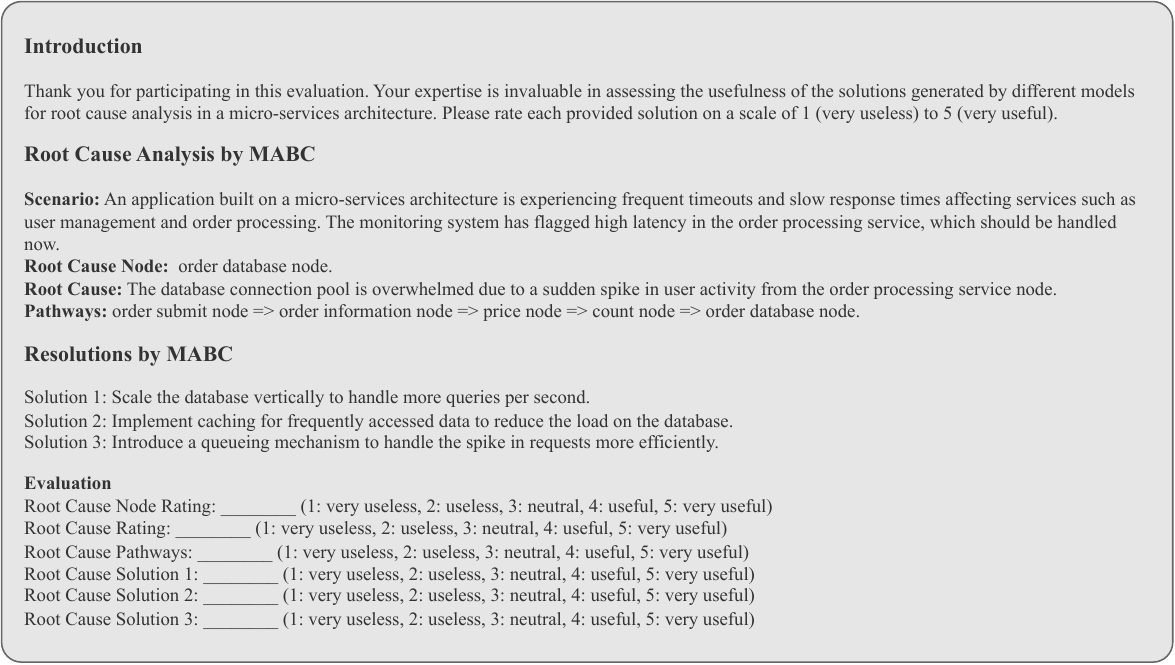}
    \caption{Human Evaluation Case.}
    \label{fig: promptE1}
\end{figure}

\clearpage
\section{Tools for Multi-Agent} \label{tool}

\begin{table}[h!]
\centering
\resizebox{2.0\linewidth}{!}{
\begin{tabular}{|>{\raggedright}p{3.5cm}|>{\raggedright}p{4.5cm}|>{\raggedright\arraybackslash}p{7cm}|}
\hline
\textbf{Agent} & \textbf{Tool Name} & \textbf{Description} \\
\hline
\textbf{Alert Receive} & Receive Alert Tool & Receive an alert from the micro-services system and add it to the scheduled queue \\
\cline{2-3}
& Prioritize Highest Alert Tool & Choose the alert with the highest priority from the scheduled queue based on the trigger time, urgency provided by self, and the number of impact nodes \\
\hline
\textbf{Process Scheduler} & Call For Help Tool & Call for help from other agents by asking a domain question and get a summary of the answer but not a detailed answer efficiently \\
\cline{2-3}
& Judge Sub-task Tool & Divide the task and conquer it by anything until no sub-task to do \\
\hline
\textbf{Data Detective} & Data Collection Tool & Collects data from nodes within a specific time window \\
\cline{2-3}
& Data Cleaning Tool & Cleans and preprocesses collected data for analysis by providing useful information but not a list of data \\
\hline
\textbf{Dependency Explorer} & Dependency Query Tool & Identifies direct and indirect dependencies of a node \\
\cline{2-3}
& Dependency Visualization Tool & Visualizes the dependencies among nodes for caching the result \\
\hline
\textbf{Probability Oracle} & Fault Probability Tool & Evaluates fault probabilities of nodes \\
\cline{2-3}
& Correlation Analysis Tool & Analyzes correlation between performance metrics of different nodes \\
\hline
\textbf{Fault Mapper} & Fault Web Tool & Visualizes and updates the fault web based on fault probabilities \\
\cline{2-3}
& Impact Analysis Tool & Analyze the impact of faults on different parts of the system \\
\hline
\textbf{Solution Engineer} & Solution Development Tool & Develops resolutions based on root cause analysis \\
\cline{2-3}
& Case Reference Tool & References previous successful cases to guide current solution development \\
\hline
\textbf{General Tools} & Metric Explorer & Retrieves node statistics like CPU and disk I/O and memory and so on for a specific time or over a time range \\
\cline{2-3}
& Alert Aggregation Tool & Aggregates alerts from various sources for unified processing \\
\cline{2-3}
& JSON Tool & Provide JSON file reading and writing \\
\hline
\end{tabular}
}
\caption{Overview of the tools for agents.}
\label{table: tools}
\end{table}